\documentclass[aps,prd,reprint,superscriptaddress,showpacs,nofootinbib]{revtex4-1}

\usepackage{graphicx}
\usepackage{dcolumn}
\usepackage{bm}
\usepackage{amsmath, amsfonts, amssymb, mathtools}

\usepackage{booktabs}
\usepackage{orcidlink}
\usepackage{subcaption}
\usepackage{xcolor}
\usepackage{listings}
\usepackage{subcaption}
\usepackage{xcolor}
\usepackage{natbib}
\usepackage{caption}
\usepackage{physics}
\usepackage{aas_macros}

\newcommand{\kfive}{\frac{5}{\ln 10}}

\numberwithin{equation}{section}

\begin{document}
\title{Testing the cosmic distance-duality relation with localized fast radio bursts: a cosmological model-independent study}

\author{Jéferson A. S. Fortunato\orcidlink{0000-0001-7983-1891}}
\email{jeferson.fortunato@uct.ac.za}
\affiliation{High Energy Physics, Cosmology \& Astrophysics Theory (HEPCAT) Group, Department of Mathematics and Applied Mathematics, University of Cape Town,
Cape Town 7700, South Africa}
\affiliation{African Institute for Mathematical Sciences, 6 Melrose Road, Muizenberg, Cape Town, 7945, South Africa}

\author{Surajit Kalita\orcidlink{0000-0002-3818-6037}}
\email{skalita@astrouw.edu.pl}
\affiliation{Astronomical Observatory, University of Warsaw, Al. Ujazdowskie 4, PL-00478 Warszawa, Poland}

\author{Amanda Weltman\orcidlink{0000-0002-5974-4114}}
\email{amanda.weltman@uct.ac.za}
\affiliation{High Energy Physics, Cosmology \& Astrophysics Theory (HEPCAT) Group, Department of Mathematics and Applied Mathematics, University of Cape Town,
Cape Town 7700, South Africa}
\affiliation{African Institute for Mathematical Sciences, 6 Melrose Road, Muizenberg, Cape Town, 7945, South Africa}

\begin{abstract}
We test the Etherington cosmic distance--duality relation (CDDR), by comparing Type Ia supernova (SN\,Ia) luminosity-distance information from the Pantheon+ compilation with an angular-diameter-distance reconstructed from localized Fast Radio Bursts (FRBs). The core of our methodology is a data-driven reconstruction from FRBs using artificial neural networks (ANNs): we infer a smooth mean extragalactic dispersion-measure relation and use its redshift derivative to recover $H(z)$ and hence $D_\mathrm{A}^{\rm FRB}(z)$ without assuming a parametric form for the expansion history. Possible deviations from CDDR are parameterized through three one-parameter models of $\eta(z)\equiv D_\mathrm{L}/[(1+z)^2D_\mathrm{A}]$. We implement two complementary likelihoods: (i) a direct approach using individual SN\,Ia with the full Pantheon+ covariance, and (ii) a machine-learning approach in which we reconstruct the SN Hubble diagram on the FRB redshift grid, propagating SN and FRB uncertainties into non-diagonal covariance matrices via Monte Carlo and bootstrap realizations. Within the FRB reconstruction, we anchor the mean extragalactic dispersion measure at $z=0$, which yields a data-driven constraint on the average host/near-source contribution $\mathrm{DM}_{\rm host}=128.8\pm 34.1\,\mathrm{pc\,cm^{-3}}$ ($3\sigma$ of statistical confidence). We find that both likelihood implementations give consistent posteriors and no statistically significant evidence for departures from CDDR at the current precision.
\end{abstract}

\maketitle

\section{Introduction}

As the name suggests, Fast Radio Bursts (FRBs) are short lived, bright radio transient events. They are characterized by narrow pulse widths lasting on the order of millisecond durations with high flux densities, typically $\mathcal{O}(\rm Jy)$. Precisely localized FRBs have emerged as powerful tools in modern cosmology. Their precise localization including redshift information makes them excellent tracers of diffuse ionized plasma that is otherwise difficult to probe. For FRBs with known redshifts, the observed dispersion measure (DM) provides us with a direct tool to understand the integrated column density of free electrons along the line-of-sight. This allows us to derive empirical constraints on the DM–redshift relation, which can be compared with theoretical models of the distribution of ionized baryons in the intergalactic medium (IGM). To date, the highest redshift of a localized FRB has been identified at redshift $z=2.148$ \cite{2025arXiv250801648C}, placing FRBs as one of the few astrophysical probes to test cosmology at relatively high redshifts. In addition, host galaxy characterization provides insight into the environments and progenitors of FRBs, reducing astrophysical uncertainties that would otherwise bias cosmological measurements.

Most FRBs are extragalactic, as indicated by their large observed DMs, hence they have become valuable probes of cosmology, including estimates of the Hubble constant $H_0$. With the increasing number of FRBs localized in recent years, the resulting cosmological constraints have steadily improved. Using a sample of eight localized FRBs, Macquart et al. \cite{macquart2020census} constrained the baryonic matter density $\Omega_\mathrm{b} = 0.051^{+0.021}_{-0.025}h_{70}^{-1}$, where $h_{70} = H_0/(70 \rm\,km\,s^{-1}\,Mpc^{-1})$, in agreement with previous estimates derived from cosmic microwave background and Big Bang nucleosynthesis data. Building on this approach, Hagstotz et al. \cite{2022MNRAS.511..662H} analyzed nine localized FRBs and inferred $H_0 = 62.3\pm 9.1 \rm\,km\,s^{-1}\,Mpc^{-1}$, assuming a homogeneous DM contribution from the host galaxies. Wu et al. \cite{2022MNRAS.515L...1W} subsequently classified 18 localized FRBs by host galaxy morphology using the IllustrisTNG simulation to estimate individual host contributions, obtaining $H_0 = 68.81^{+4.99}_{-4.33}\rm\,km\,s^{-1}\,Mpc^{-1}$. James et al. \cite{2022MNRAS.516.4862J} expanded the analysis to include 16 localized and 60 unlocalized FRBs detected by ASKAP, obtaining $H_0 = 73^{+12}_{-8}\rm\,km\,s^{-1}\,Mpc^{-1}$. Since then, a range of studies has explored alternative statistical techniques, simulations, and FRB samples to further refine constraints on $H_0$ \cite{2024ApJ...965...57B,2023ApJ...946L..49L,2023ApJ...955..101W,2024MNRAS.527.7861G,kalita2025fast,2026ApJ...996...66Z}. More recently, introduction of a machine-learning approach based on artificial neural networks (ANNs) have emerged as an alternate way to estimate $H_0$ and thereby test the underlying cosmology \cite{fortunato2025fast, fortunato2025probing, kalita2025revealing}.

The Etherington cosmic distance-duality relation (CDDR) is a fundamental identity in metric theories of gravity \cite{etherington1933lx}, which connects two distance measures in cosmology: the luminosity distance $D_\mathrm{L}$ and the angular diameter distance $D_\mathrm{A}$. It is defined as
\begin{equation}
    \eta(z) \equiv \frac{D_\mathrm{L}(z)}{(1 + z)^2 D_\mathrm{A}(z)} = 1.
\end{equation}
This CDDR holds in any metric theory of gravity, provided that photons propagate along null geodesics and their number is conserved. Under these conditions, any of the above two cosmological distances can be used to make measurements as they are dual to each other. Importantly, this relation is independent of the cosmological model and the dynamics of cosmic expansion, making it a powerful consistency condition. Testing this relation with independent observables provides a direct way to probe violations of photon conservation, exotic physics, or cosmic opacity.

A large body of work has tested the CDDR by pairing $D_\mathrm{L}$ indicators with independent determinations of $D_\mathrm{A}$ \cite{2004PhRvD..69j1305B,2017Univ....3...52S,2023MNRAS.518.5490G}. Classic examples include combining Type Ia supernovae (SNe\,Ia) with galaxy-cluster distances inferred from joint Sunyaev--Zel'dovich (SZ) and X-ray observations \cite{uzan2004distance}, as well as comparisons involving Baryon acoustic oscillations (BAO) measurements, gravitational waves and SNe\,Ia \cite{de2025distinguishing}. In most analyses, $\eta(z)$ remains consistent with unity within current uncertainties, and apparent departures are typically attributed to residual systematics or to effective photon non-conservation mechanisms such as cosmic opacity or photon--axion mixing \cite{xu2020probing, avgoustidis2010constraints}. In this context, CDDR tests are often framed as null experiments: any statistically significant detection of $\eta(z)\neq 1$ would point either to unaccounted observational biases or to new physics affecting photon propagation.

From a phenomenological standpoint, CDDR deviations are commonly modeled through low-dimensional parametrizations of $\eta(z)$ (or equivalently, an opacity function in magnitude space), and constraints are obtained by fitting $\eta(z)$ jointly with nuisance parameters that absorb distance-scale calibration \cite{gahlaut2025model, nair2012cosmic}. Importantly, enforcing $\eta(0)=1$ is not an arbitrary choice: a redshift-independent offset in $\eta$ is exactly degenerate with an additive magnitude calibration (e.g.\ the standardized SN absolute magnitude), so only the \emph{redshift dependence} of $\eta(z)$ is empirically identifiable in distance-ladder--free analyses. This motivates parametrizations that reduce to unity at $z\to 0$ and target differential deviations across redshift.

FRBs provide a complementary route for the CDDR test because their primary cosmological observable, the DM, encodes the integrated free-electron column density along the line of sight and is therefore sensitive to geometry through the reconstructed expansion history, rather than relying on photon fluxes. Once a smooth mapping between extragalactic DM and $z$ is reconstructed, one can infer $H(z)$ in a minimally model-dependent way and thereby obtain $D_\mathrm{C}(z)$ and $D_\mathrm{A}(z)$. This makes FRB-based distances particularly attractive for CDDR studies: they are subject to very different astrophysical and instrumental systematics than standard candles.

In this work, we combine FRB-reconstructed angular diameter distances, $D_\mathrm{A}^{\rm FRB}(z)$, with SN\,Ia luminosity distance information from the Pantheon+ compilation. We implement the comparison in a statistically consistent framework that explicitly propagates the \emph{correlated} uncertainties induced by global reconstructions. The FRB distance curve inherits non-trivial covariance from bootstrap/ensemble reconstructions, and (in one of our implementations) the SN Hubble diagram is also reconstructed on the FRB redshift grid while carrying the full Pantheon+ covariance into a non-diagonal prediction covariance. 

The article is organized as follows. In Sec.~\ref{Sec2}, we review the basic properties of the DM of FRBs, and in Sec.~\ref{Sec3} we summarize the cosmological distance measures relevant to this work. Sec.~\ref{sec:dm_to_da} describes the algorithm used to estimate the mean DM contributions from the host galaxy and IGM components with our machine-learning (ML) framework. In Sec.~\ref{Sec5}, we present the sample of localized FRBs and, using these data, reconstruct the mean DM--redshift relation. Sec.~\ref{sec:likelihood_method} outlines the full methodology to test the CDDR, while Sec.~\ref{Sec7} presents the resulting posterior constraints on the function $\eta(z)$. Finally, in Sec.~\ref{Sec8}, we summarize our findings and provide concluding remarks.

\section{The Dispersion Measure from FRBs}\label{Sec2}

The observed DM of a localized FRB can be decomposed as
\begin{equation}
\mathrm{DM}_{\rm obs}=\mathrm{DM}_{\rm MW}+\mathrm{DM}_{\rm halo}+\mathrm{DM}_{\rm IGM}(z)+\frac{\mathrm{DM}_{\text{host}}}{1 + z},
\end{equation}
where each term traces a distinct electron column density along the line-of-sight. The Milky Way (MW) contribution, $\mathrm{DM}_{\rm MW}$, arises primarily from the Galactic interstellar medium and is commonly estimated using Galactic free-electron density models such as NE2001 \cite{cordes2002ne2001} or YMW16 \cite{yao2017new}. In this work, we adopt NE2001, motivated by indications that YMW16 can overestimate $\mathrm{DM}_{\rm MW}$ in some low-latitude directions \cite{koch2021}. The Galactic halo contribution, $\mathrm{DM}_{\rm halo}$, is less well constrained; representative values of $\sim 50-100\,\mathrm{pc}\,\mathrm{cm}^{-3}$ have been inferred from a combination of observational and dynamical arguments, including DM of the Large Magellanic Cloud, high-velocity cloud constraints, and hydrostatic models of the hot circumgalactic medium \cite{prochaska2019probing}. To remain conservative, we fix $\mathrm{DM}_{\rm halo}=50\,\mathrm{pc}\,\mathrm{cm}^{-3}$ throughout.

The intergalactic medium contribution, $\mathrm{DM}_{\mathrm{IGM}}$, provides the dominant cosmological signal in the DM of localized FRBs. While individual sightlines can deviate substantially due to large-scale structure and inhomogeneities in the free-electron distribution, the ensemble-averaged relation admits a well-defined mean. Following \cite{deng2014cosmological}, the expectation value is
\begin{equation}\label{DM_igm}
\langle \mathrm{DM}_{\mathrm{IGM}}(z) \rangle =
\left(\frac{3c \Omega_\mathrm{b} H_0}{8\pi G m_\mathrm{p} }\right)
\int_0^z \frac{(1+z')\, f_{\mathrm{IGM}}(z')\, f_\mathrm{e}(z')}{E(z')}\dd{z}',
\end{equation}
where $c$ is the speed of light, $G$ is the gravitational constant, and $m_\mathrm{p}$ is the proton mass. Here $\Omega_\mathrm{b}$ denotes the baryon density parameter, and $E(z)= H(z)/H_0$ with $H(z)$ being the Hubble function. The factor $f_{\mathrm{IGM}}(z)$ represents the fraction of baryons residing in the diffuse IGM. We adopt a redshift-independent value $f_{\mathrm{IGM}}=0.82\pm0.04$ \cite{shull2012baryon, zhou2014fast}. The term $f_\mathrm{e}(z)$ encodes the ionization state, and assuming only hydrogen and helium are present, it is given by
\begin{equation}\label{ioniza}
f_\mathrm{e}(z)=Y_\mathrm{H}\,X_{\mathrm{e},\mathrm{H}}(z)+\frac{1}{2}Y_{\mathrm{He}}\,X_{\mathrm{e},\mathrm{He}}(z),
\end{equation}
where $Y_\mathrm{H}$ and $Y_{\mathrm{He}}$ are the hydrogen and helium mass fractions, and $X_{\mathrm{e},\mathrm{H}}$ and $X_{\mathrm{e},\mathrm{He}}$ are their ionization fractions.


The host-galaxy contribution, $\mathrm{DM}_{\mathrm{host}}$, depends on the local environment of the burst (e.g.\ star-forming regions, SN remnants, or circumburst plasma) and is therefore among the least well-constrained terms in the DM budget. In the observer frame, it is diluted by cosmic expansion, entering in the total DM as $\mathrm{DM}_{\mathrm{host}}/(1+z)$. A common simplifying assumption is to adopt a constant host term, e.g. $\mathrm{DM}_{\mathrm{host}}\approx 100\,\mathrm{pc}\,\mathrm{cm}^{-3}$ \cite{tendulkar2017host,2022PhRvD.106d3017L,2023JCAP...11..059K}; however, for low-redshift FRBs this can lead to unphysical inferences, including negative $\mathrm{DM}_{\mathrm{IGM}}$ after subtraction. More flexible alternatives have been proposed, such as redshift-dependent prescriptions calibrated on simulations (e.g.\ IllustrisTNG) \cite{zhang2020dispersion}. In this work, we follow the conservative strategy of \cite{kalita2025revealing} and avoid imposing an explicit parametric model for $\mathrm{DM}_{\mathrm{host}}(z)$. Instead, we use it as an input to our reconstruction pipeline of the extragalactic DM,
\begin{eqnarray}
    \mathrm{DM}_{\text{EG}}(z) &=& \mathrm{DM}_{\text{obs}} - \mathrm{DM}_{\text{MW}} - \mathrm{DM}_{\text{halo}}\\
    &=& \mathrm{DM}_{\text{IGM}}(z) + \frac{\mathrm{DM}_{\text{host}}}{1 + z}.
\end{eqnarray}
This choice minimizes assumptions about the host environment while retaining the cosmological information carried by the redshift evolution of the mean extragalactic DM. The decomposition above isolates $\mathrm{DM}_{\rm EG}$ as the key observable enabling FRBs to be used as cosmological probes, as developed in the following sections.

\section{Distances in Cosmology}\label{Sec3}

The line-of-sight comoving distance $D_\mathrm{C}$ provides a convenient, expansion-independent measure of separation in an FLRW spacetime. It is obtained directly from the expansion history as
\begin{equation}\label{eq:comoving_distance}
D_\mathrm{C}(z)=c\int_0^z \frac{\dd{z}'}{H(z')}.
\end{equation}
It is often useful to introduce the Hubble distance, defined locally by
\begin{equation}
D_H(z)\equiv \frac{c}{H(z)},
\end{equation}
with the present-day value $D_{H_0}=c/H_0$. This quantity sets a characteristic length scale associated with the instantaneous expansion rate and should not be conflated with particle or event horizons, which involve time integrals of $H(z)$.

Spatial curvature modifies the relation between the radial comoving distance and the transverse comoving distance (also called the ``comoving angular diameter'' distance). Denoting the curvature parameter by $\Omega_k$, the transverse distance can be written as \citep{weinberg1972gravitation}
\begin{equation}\label{eq:DM}
D_\mathrm{M}(z)=
\begin{cases}
\dfrac{D_{H_0}}{\sqrt{\Omega_k}}\,
\sinh\!\left(\sqrt{\Omega_k}\,\dfrac{D_\mathrm{C}(z)}{D_{H_0}}\right), & \Omega_k>0,\\[8pt]
D_\mathrm{C}(z), & \Omega_k=0,\\[8pt]
\dfrac{D_{H_0}}{\sqrt{|\Omega_k|}}\,
\sin\!\left(\sqrt{|\Omega_k|}\,\dfrac{D_\mathrm{C}(z)}{D_{H_0}}\right), & \Omega_k<0,
\end{cases}
\end{equation}
corresponding to open, flat, and closed geometries, respectively. Observationally, distances are typically accessed through the angular diameter distance, which is related to the transverse comoving distance by the usual redshift factor,
\begin{eqnarray}\label{DAeq}
D_\mathrm{A}(z)&=&\frac{D_\mathrm{M}(z)}{1+z}\equiv \frac{D_\mathrm{C}(z)}{1+z}.
\end{eqnarray}
In this study, we restrict our analysis to the spatially flat geometry. A detailed discussion of how spatial curvature may impact the present methodology can be found in \cite{fortunato2025probing}. Nevertheless, these relations provide a geometric link between the reconstructed $H(z)$ from FRBs and the inferred $D_\mathrm{A}^{\rm FRB}(z)$ used in our CDDR analysis.

\section{From Dispersion Measure to Angular Diameter Distance}\label{sec:dm_to_da}

The central ingredient that allows FRBs to act as cosmological probes is the redshift evolution of the extragalactic DM. Following the procedure outlined in \cite{fortunato2025probing}, Eq.~\eqref{DM_igm} can be written as
\begin{equation}
\langle \mathrm{DM}_{\mathrm{IGM}}(z)\rangle = A \int_{0}^{z} F(z')\, \dot D_\mathrm{C}(z')\dd{z}',
\label{digm2}
\end{equation}
where 
\begin{equation}
    \dot D_\mathrm{C}\equiv \dv{D_\mathrm{C}}{z} = \frac{c}{H(z)},
\end{equation}
with
\begin{equation}
    A = \frac{3c \, \Omega_\mathrm{b} H_0^2}{8 \pi G m_\mathrm{p}}.
\end{equation}

The redshift-dependent modulation
\begin{equation}
    F(z)=(1+z)\, f_{\mathrm{IGM}}(z)\, f_\mathrm{e}(z)
\end{equation}
parametrizes both the fraction of baryons contained in the diffuse IGM and the ionization state of the gas (see Sec.~\ref{Sec2}). Moreover, Eq.~\eqref{digm2} explicitly shows that the cosmological dependence arises through $\dot D_\mathrm{C}$, whereas astrophysical contributions are entirely captured by $F(z)$.

Differentiating Eq.~(\ref{digm2}) with respect to redshift yields a direct relation between the slope of the mean Macquart curve and the inverse expansion rate, i.e.
\begin{equation}
\dv{\langle \mathrm{DM}_{\mathrm{IGM}}(z)\rangle}{z} = A\,F(z)\,\frac{c}{H(z)} \, ,
\end{equation}
or equivalently,
\begin{equation}\label{dotDc}
\dot D_\mathrm{C}(z)=\frac{1}{A\,F(z)}\,
\dv{\langle \mathrm{DM}_{\mathrm{IGM}}(z)\rangle}{z}\, .
\end{equation}
This expression motivates our reconstruction strategy: rather than assuming a specific parametric form for $H(z)$, we infer the $\dv*{\langle \mathrm{DM}_{\mathrm{IGM}}(z)\rangle}{z}$ relation from the data and explicitly remove the modulation $F(z)$. In this way, the expansion history enters only through the reconstructed curve, making the procedure minimally model-dependent. Once $\dot D_\mathrm{C}(z)$ is reconstructed, the comoving distance can be obtained from Eq.~\eqref{eq:comoving_distance}, and thereby the angular diameter distance from Eq.~\eqref{DAeq}.

Individual FRBs exhibit substantial scatter around the mean relation due to inhomogeneities in the free-electron distribution and uncertain host contributions. We therefore do not interpret Eq.~(\ref{digm2}) as an exact point-by-point mapping. Instead, we reconstruct the mean trend $\langle \mathrm{DM}_{\rm EG}(z)\rangle$ using an ensemble of neural networks and propagate uncertainties through bootstrap/ensemble realizations. These realizations naturally induce correlated uncertainties across redshift in the derived quantities $H(z)$, $D_\mathrm{C}(z)$, and $D_\mathrm{A}(z)$, which are encapsulated in the full covariance matrix used in the subsequent CDDR likelihood analysis.

The cosmological contribution must vanish as $z\to 0$, whereas a residual extragalactic DM at very low redshift is naturally interpreted as an effective host/near-source term. Motivated by this, we reconstruct a smooth $\langle \overline{\mathrm{DM}}_{\rm EG}(z)\rangle$ and enforce the physical condition $\langle \overline{\mathrm{DM}}_{\rm IGM}(0)\rangle=0$ by subtracting the reconstructed intercept, i.e.
\begin{equation}
 \langle{\overline{\mathrm{DM}}_{\rm IGM}(z)}\rangle\equiv  \overline{\mathrm{DM}}_{\rm EG}(z)- \overline{\mathrm{DM}}_{\rm EG}(0).
\label{eq:dm_intercept_sub}
\end{equation}
The bars mean these are the reconstructed quantities. In this effective description, $\overline{\mathrm{DM}}_{\rm EG}(0)$ plays the role of a mean host/near-source contribution in the observer frame. We emphasize that this operation is applied at the level of the reconstructed extragalactic contribution mean trend, not to individual bursts, and therefore acts as a stable anchor that mitigates unphysical negative $\mathrm{DM}_{\rm IGM}$ at low redshift.

To reconstruct $\dv*{\langle \overline{\mathrm{DM}}_{\rm IGM}\rangle}{z}$ in Eq.~(\ref{dotDc}), we follow the data-driven strategy introduced in \cite{fortunato2025fast} to the present CDDR analysis. Concretely, we train an ensemble of feed-forward multilayer perceptrons (MLPs) implemented in \texttt{scikit-learn} \cite{pedregosa2011scikit} to learn the mapping $z\mapsto \langle \overline{\mathrm{DM}}_{\rm EG}(z)\rangle$. We explore a grid of architectures and optimization settings (hidden-layer widths, activation functions, solvers, regularization strength, and learning-rate schemes) using cross-validation to select a configuration that provides a smooth fit while controlling overfitting. Once the optimal configuration is identified, we evaluate the reconstructed mean curve on a dense redshift grid and compute its derivative numerically, yielding an estimate of $\dv*{\langle \overline{\mathrm{DM}}_{\rm IGM}\rangle}{z}$ after applying the intercept correction. The intrinsic scatter and measurement uncertainties are propagated into the derived distances, as we generate an ensemble of reconstructions by bootstrap resampling of the FRB sample. For each realization $k$, we obtain a smooth curve $ {\mathrm{DM}}_{\rm EG}^{(k)}(z)$, apply the intercept anchoring, and then compute the corresponding $\dot D_\mathrm{C}^{(k)}(z)=c/H^{(k)}(z)$ using Eq.~(\ref{dotDc}) and thereby $D_\mathrm{C}^{(k)}(z)$ and $D_\mathrm{A}^{(k)}(z)$. For transparency and reproducibility, we plan to release this reconstruction pipeline as a public tool for smooth function inference with correlated uncertainty propagation in a future work.

The output of the procedure above is a cosmological model-agnostic reconstruction of the angular diameter distance, $D_\mathrm{A}^{\rm FRB}(z)$. This constitutes the $D_\mathrm{A}$ ingredient of the Etherington relation. To validate CDDR relation, we require an independent estimate of $D_\mathrm{L}(z)$. In this work, we obtain $D_\mathrm{L}$ from the SN\,Ia Pantheon+ compilation \cite{scolnic2018complete}. The two distance measures are then combined through $\eta(z)\equiv D_\mathrm{L}/[(1+z)^2D_\mathrm{A}]$ in the likelihood framework presented in Sec.~\ref{sec:likelihood_method}.

\section{Data and reconstructions}\label{Sec5}

To validate the CDDR relation, we use the localized FRBs compiled in \cite{fortunato2025probing}. With the requirement of a measured redshift and a reported DM, we retain a cleaned subset of 122 events\footnote{131 localized FRBs were reported in the catalog as of November 2025}. We additionally impose a conservative quality cut to ensure a physically meaningful extragalactic DM budget and to stabilize the subsequent distance reconstruction. Following the practical criterion adopted in \cite{wang2025probing}, we require
\begin{equation}
\mathrm{DM}_{\rm obs}-\mathrm{DM}_{\rm MW} > 80\,\mathrm{pc\,cm^{-3}},
\end{equation}
which effectively accounts for $\mathrm{DM}_{\rm halo}\sim 50-80\,\mathrm{pc\,cm^{-3}}$. This condition acts as a proxy for demanding a positive extragalactic DM, i.e. $\mathrm{DM}_{\rm EG} > 0$; thereby avoiding unphysical negative values when isolating the cosmological component. Beyond basic physical consistency, excluding very low-$\mathrm{DM}_{\rm EG}$ events is crucial because they can disproportionately influence a flexible reconstruction of $\langle \mathrm{DM}(z)\rangle$, biasing the inferred distance scale. In the context of our CDDR test, this would translate into a biased $D_\mathrm{A}^{\rm FRB}(z)$, potentially inducing spurious apparent deviations in the considered parametrizations. The cut also tends to remove part of the lowest-redshift tail, where peculiar velocities and local-structure fluctuations can be non-negligible relative to the cosmological signal.

A priori we remove a small number of events for specific, well-motivated reasons. FRB\,200110E is excluded because it lies at an extremely small distance ($\sim 3.6\,\mathrm{Mpc}$) within a globular cluster in the M81 group \citep{kirsten2022repeating, bhardwaj2021nearby}. At such proximity, the IGM contribution is negligible, making the burst unsuitable for cosmological inference. We also exclude FRB\,210117 and FRB\,190520, whose measured DMs are anomalously large for their redshifts. These sources are commonly treated as DM outliers, plausibly dominated by dense or highly magnetized local environments in their host systems \citep{bhandari2023nonrepeating, gordon2023demographics, niu2022repeating}, and would otherwise distort a smooth mean reconstruction.

Finally, we note an intrinsic limitation of current localized samples: the number of FRBs becomes sparse at $z\gtrsim 1$. While our baseline analysis includes all events passing the cleaning steps above, only a handful of FRBs populate the high-$z$ tail (e.g. FRB\,220610, FRB\,221029, FRB\,230521, FRB\,240123, and the most distant event FRB\,240304B at $z=2.148$). In this regime, any flexible reconstruction method (including our ANN ensemble) is weakly anchored by the data leading to broadening of the predictive variance. This behavior is expected and is consistently propagated into the uncertainties of $D_\mathrm{A}^{\rm FRB}(z)$ and the resulting CDDR constraints through our bootstrap/ensemble procedure.

\begin{figure}[htp]
  \centering

  \includegraphics[width=\linewidth]{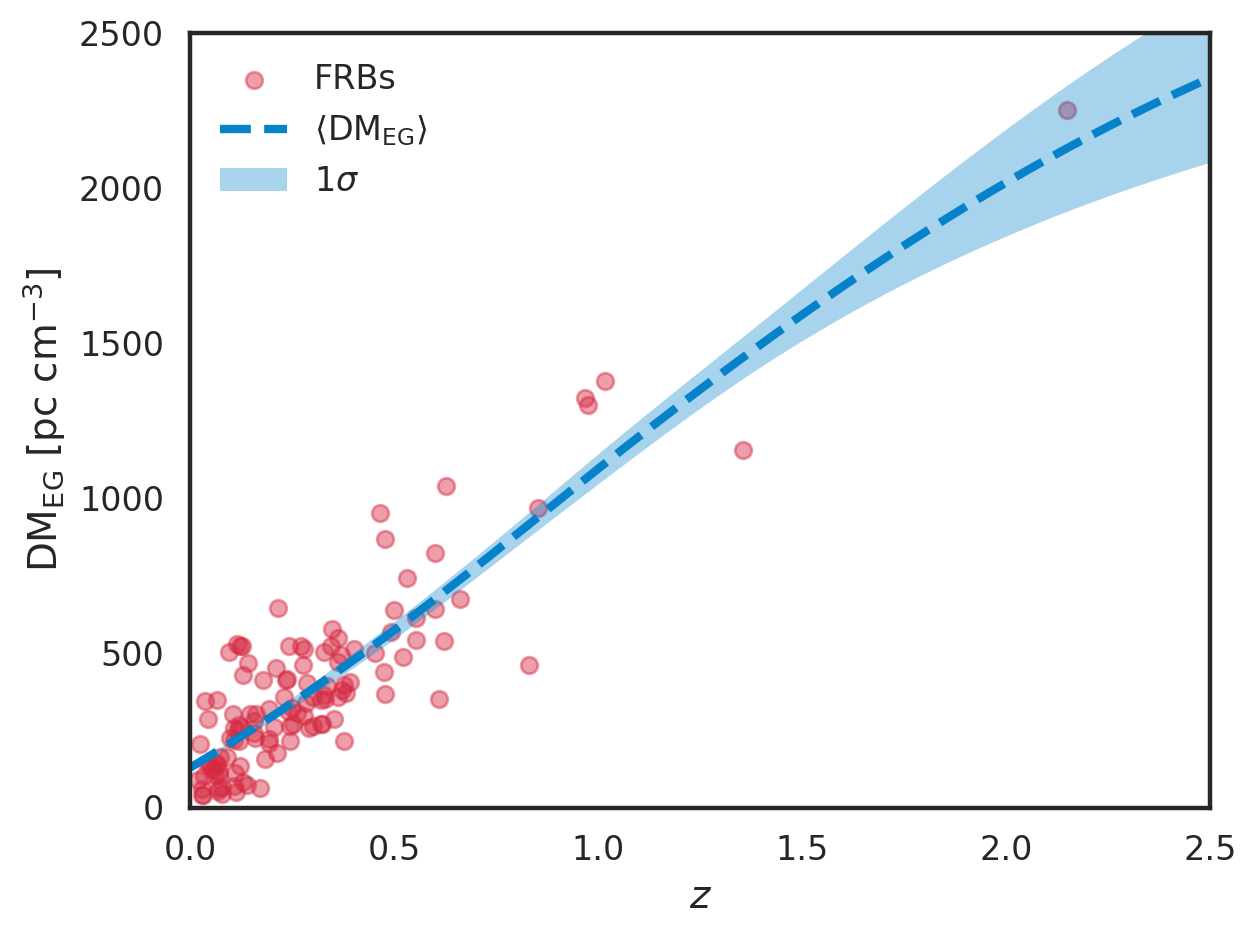}\\[-2pt]
  \caption*{\textbf{(a)} $\mathrm{DM}_{\mathrm{EG}}(z)$ reconstruction.}

  \vspace{4pt}
  \includegraphics[width=\linewidth]{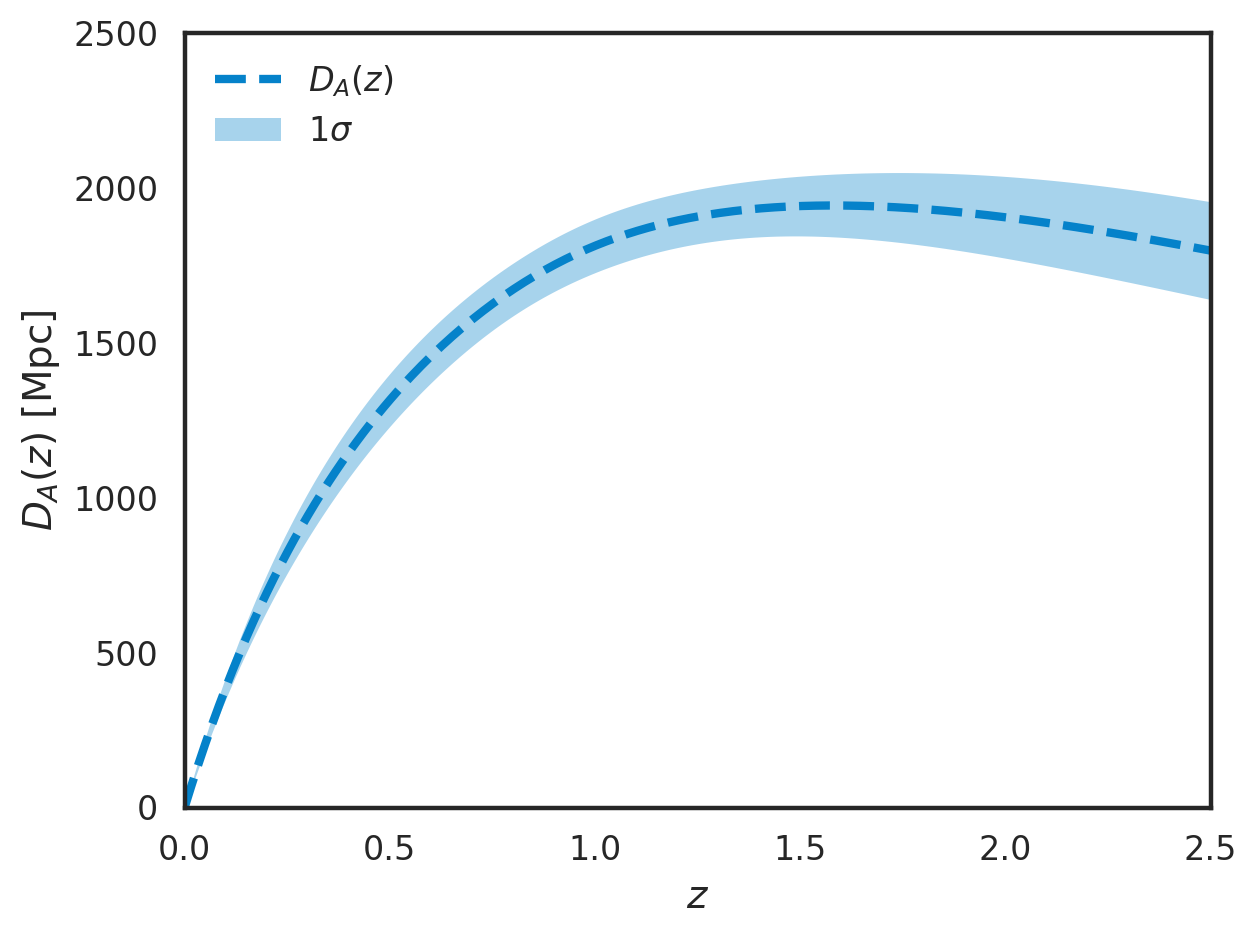}\\[-2pt]
  \caption*{\textbf{(b)} Reconstructed $D_\mathrm{A}^{\rm FRB}(z)$.}

  \vspace{4pt}
  \includegraphics[width=\linewidth]{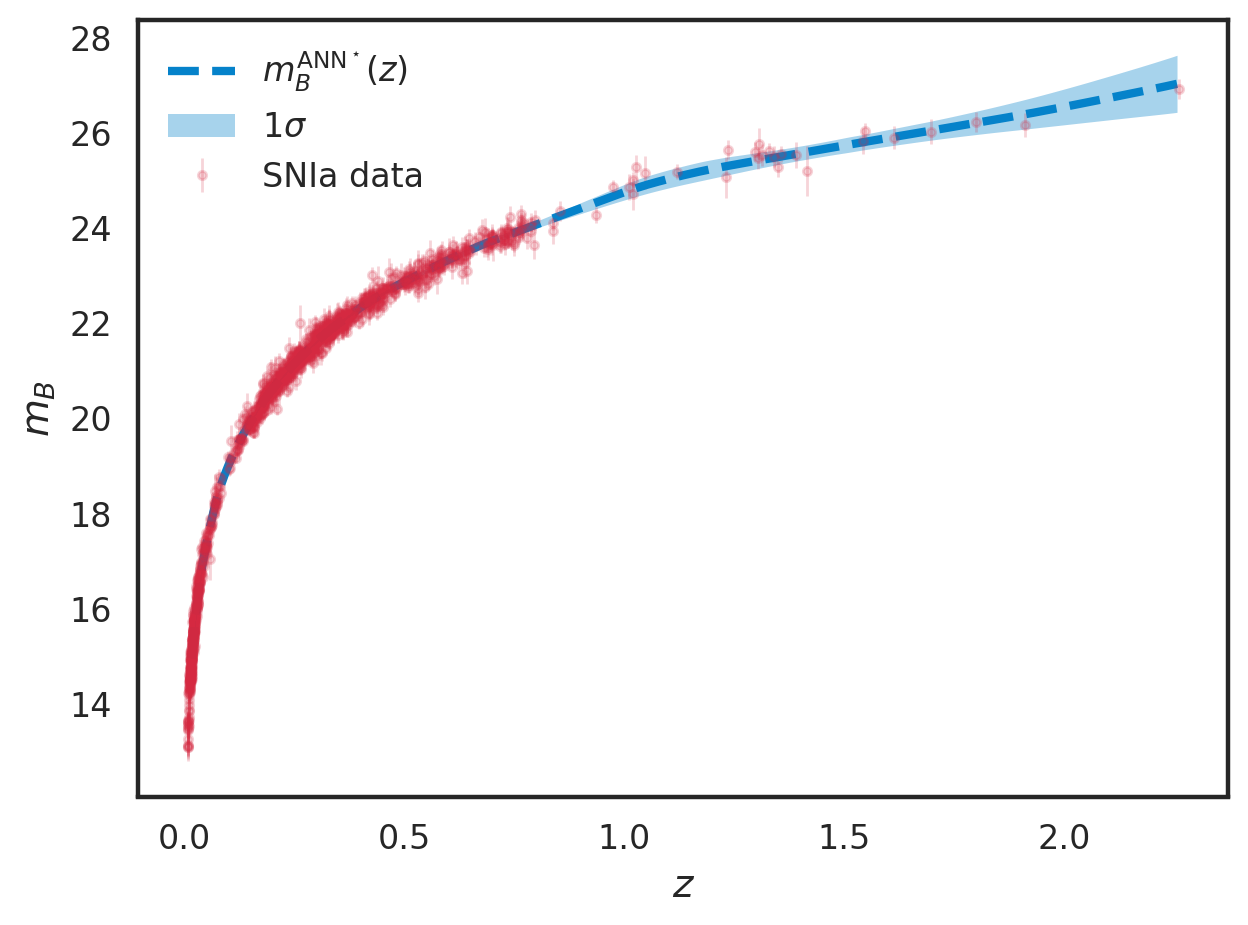}\\[-2pt]
  \caption*{\textbf{(c)} Reconstructed apparent magnitude from SN\,Ia data.}

  \caption{Reconstructions used in the pipeline. Panels (a) and (b) show the FRB-based reconstructions of $\rm DM_{\rm EG}(z)$ and $D_\mathrm{A}^{\rm FRB}(z)$, while panel (c) shows the ANN reconstruction of the SN Hubble diagram used in Method~B.}
  \label{fig:recs_stack}
\end{figure}

To apply the aforementioned methodology, we start from the cleaned subset of 122 localized FRBs. For each event, we remove the DM contributions from the MW and the halo from the observed DM to obtain the extragalactic contribution. This yields an effective mean $ {\mathrm{DM}}_{\rm IGM}(z)$ that is suitable for derivative-based inference of the expansion history. Fig.~\ref{fig:recs_stack}(a) shows ${\mathrm{DM}}_{\rm EG}$ vs. $z$ for each FRB (red dots) together with the ANN reconstruction (solid blue curve) alongside the $1\sigma$ uncertainty shown in a shaded band derived from the bootstrap/ensemble procedure. 
Using the methodology discussed in Sec.~\ref{sec:dm_to_da}, the resulting $D_\mathrm{A}^{\rm FRB}(z)$ reconstruction is shown in Fig.~\ref{fig:recs_stack}(b), along with $1\sigma$ uncertainties. For our cleaned FRB sample, we obtain $\mathrm{DM}_{\mathrm{host}}\equiv \overline{\mathrm{DM}}_{\rm EG}(0)=128.77\pm 11.37\,{\rm pc\,cm^{-3}} (1\sigma)$ and with a conservative $3\sigma$ interval, it corresponds
$\mathrm{DM}_{\mathrm{host}}\equiv \overline{\mathrm{DM}}_{\rm EG}(0)\in[94.67,\,162.87]\,{\rm pc\,cm^{-3}}$.
In this effective description, this component plays the role of a mean host/near-source contribution (including any local circumburst component) for the ensemble, and its magnitude is broadly consistent with the $\mathcal{O}(10^2)\,{\rm pc\,cm^{-3}}$ host terms commonly adopted or inferred in the FRB cosmology literature (e.g.\ constant-host prescriptions and simulation-calibrated estimates; see, e.g., \cite{tendulkar2017host, zhang2020dispersion}).

For the $D_\mathrm{L}$ side of the CDDR test, we use the Pantheon+ compilation of SN\,Ia \cite{scolnic2018complete}, adopting the publicly released corrected apparent magnitudes $m_\mathrm{B}$ (equivalently, distance moduli up to an absolute-magnitude offset) together with the full statistical+systematic covariance matrix. To avoid introducing external distance anchors, we remove the SH0ES calibrator subset and any objects flagged as used in the SH0ES Hubble-flow analysis, following the metadata provided in the release. We use the catalogue as an input for the ANN reconstruction the apparent magnitude and its corresponding redshift, depicted in Fig.~\ref{fig:recs_stack}(c).

\section{Cosmic distance-duality test: likelihood and methodology}\label{sec:likelihood_method}

Our goal is to test the Etherington relation by combining $D_\mathrm{A}^{\rm FRB}(z)$ with $D_\mathrm{L}$ from Pantheon+ SNe\,Ia sample. In magnitude space, the SN prediction can be written as
\begin{equation}
m_\mathrm{B}(z)=\mu(z)+M_\mathrm{B},
\end{equation}
where $M_\mathrm{B}$ is the standardized absolute magnitude and
\begin{equation}
\mu(z)=5\log_{10}\!\left[\frac{D_\mathrm{L}(z)}{\mathrm{Mpc}}\right]+25
\end{equation}
is the distance modulus. Since $M_\mathrm{B}$ absorbs the overall calibration of the SN distance ladder, we treat it as a free nuisance parameter in all fits and do not impose external anchors.

In our CDDR analysis, we combine SN information with the FRB-based angular diameter distance by writing the luminosity distance as
\begin{equation}
D_\mathrm{L}^{\rm FRB}(z)=(1+z)^2\,D_\mathrm{A}^{\rm FRB}(z)\,\eta(z),
\end{equation}
so that the model in magnitude space becomes
\begin{equation}
m_\mathrm{B}^{\rm FRB}(z)=\mu_0(z)+\Delta_\eta(z)+M_\mathrm{B},
\end{equation}
with $\mu_0(z)=25+5\log_{10}\!\big[(1+z)^2D_\mathrm{A}^{\rm FRB}(z)\big]$ and $\Delta_\eta(z)=5\log_{10}\eta(z)$. 
Possible deviations from CDDR are encoded through
\begin{equation}
\eta(z)\equiv \frac{D_\mathrm{L}^{\rm SN}(z)}{(1+z)^2D_\mathrm{A}^{\rm FRB}(z)}\,,
\end{equation}
which enters additively as
\begin{equation}
\Delta_\eta(z)\equiv 5\log_{10}\eta(z)=\kfive\,\ln\eta(z).
\label{eq:delta_eta}
\end{equation}
We treat the standardized SN $M_\mathrm{B}$ as a free parameter.

A subtle but important aspect of this analysis is the treatment of uncertainties. Individual SN and FRB measurements are independent, but both $D_\mathrm{A}^{\rm FRB}(z)$ and (in one of our approaches) the SN Hubble diagram are obtained through global reconstructions, which naturally induce correlated uncertainties across redshift. We therefore retain the full covariance of the reconstructed quantities in the likelihood.

We work in magnitude space using the Pantheon+ corrected apparent magnitudes $m_\mathrm{B}^{\rm SN}$
and the FRB-based $m_\mathrm{B}^{\rm FRB}$.
Defining the residual vector
\begin{equation}\label{residualvec}
\mathbf{r}(\theta,M_\mathrm{B})\equiv \mathbf{m}_\mathrm{B}^{\rm SN}-\mathbf{m}_\mathrm{B}^{\rm FRB}(\theta,M_\mathrm{B}),
\end{equation}
we assume a multivariate Gaussian likelihood,
\begin{equation}
-2\ln\mathcal{L} = \mathbf{r}^{\top} C^{-1}\mathbf{r}.
\label{eq:gauss_like_simple}
\end{equation}
The total covariance matrix $C$ depends on how the SN information is incorporated (Method~A or Method~B), as described below. 

\begin{table*}[htpb]
    \centering
    \caption{Posterior means $\pm1\sigma$ for the CDDR deviation parameter and $M_\mathrm{B}$, comparing Method A (FULL) and Method B (ANN).}
    \begin{tabular}{| l | c | c | c | c |}
        \hline
        Model & $\theta_{\rm FULL}$ & $M_{B,{\rm FULL}}$ & $\theta_{\rm ANN}$ & $M_{B,{\rm ANN}}$ \\
        \hline
        linear  & $\epsilon = -0.0221 \pm 0.0482 $ & $ -19.5047 \pm 0.1446$ & $\epsilon = -0.0263 \pm 0.0628$ & $-19.5448 \pm 0.1552$ \\
        linfrac & $\epsilon = -0.0074 \pm 0.1116$ & $-19.5508 \pm 0.1445$ & $\epsilon = 0.0891 \pm 0.1624$ & $-19.6683 \pm 0.1710$ \\
        power   & $\beta   = -0.0355 \pm 0.0837$ & $-19.5110 \pm 0.1532$ & $\beta = -0.0189 \pm 0.1073$ & $-19.5710 \pm 0.1687$  \\ 
        \hline
    \end{tabular}
    \label{tab:main}
\end{table*}

\subsection{Method A: direct Pantheon+ likelihood (FULL)}
\label{sec:methodA_new}

In Method A, we work directly with the Pantheon+ corrected apparent magnitudes. Let $\{z_i\}$ be the redshifts of the selected SN -- within the FRBs redshift range -- used in the comparison. For each SN redshift, we build the FRB-based model prediction by interpolating $D_\mathrm{A}^{\rm FRB}(z)$ to $z_i$ and thereby writing $m_{\mathrm{B},i}^{\rm FRB}$. The residual vector, given by Eq.~\eqref{residualvec}, is therefore evaluated at the SN redshifts. The SN uncertainties are described by the Pantheon+ sample statistical+systematic covariance submatrix $C_{\rm SN,sub}$ restricted to the same SN subset. Uncertainties in the FRB reconstruction (including the induced correlations across redshift) are propagated to magnitude space and summarized by a covariance matrix $C_{\rm FRB}$ evaluated at the same SN redshifts. The total covariance is
\begin{equation}
C_\mathrm{A} \equiv C_{\rm SN,sub}+C_{\rm FRB}.
\end{equation}
Eq.~(\ref{eq:gauss_like_simple}) with $\mathbf{r}=\mathbf{r}_\mathrm{A}$ and $C=C_\mathrm{A}$
defines the FULL likelihood.

\subsection{Method B: reconstructed SN Hubble diagram on the FRB grid (ANN)}
\label{sec:methodB_new}

Method B compresses the SN information into a smooth reconstruction evaluated on the FRB redshift grid, enabling a direct comparison with the FRB-based baseline. Let $\{z_j\}$ denote the redshift grid on which the FRB reconstruction provides $D_\mathrm{A}^{\rm FRB}(z_j)$ (and hence $\mu_0(z_j)$). We reconstruct a smooth SN curve $m_\mathrm{B}^{\rm ANN}(z)$ on the same grid while propagating the \emph{full} Pantheon+ covariance into a non-diagonal prediction covariance.

Concretely, starting from the Pantheon+ data vector $\mathbf{m}_\mathrm{B}^{\rm SN}$ and its full covariance $C_{\rm SN}$, we generate parametric Monte Carlo realizations
\begin{equation}
\mathbf{m}_\mathrm{B}^{(k)} \sim \mathcal{N}\!\left(\mathbf{m}_\mathrm{B}^{\rm SN},\, C_{\rm SN}\right),
\end{equation}
train one neural network per realization to learn $m_\mathrm{B}(z)$, and evaluate each trained network on the FRB grid $\{z_j\}$. The ensemble of network predictions defines the mean reconstructed SN curve,
\begin{equation}
\bar m_\mathrm{B}^{\rm ANN}(z_j) \equiv \left\langle m_\mathrm{B}^{(k)}(z_j)\right\rangle,
\end{equation}
and the corresponding prediction covariance,
\begin{equation}
(C_{\rm ANN})_{ij}\equiv
\left\langle\big[m_\mathrm{B}^{(k)}(z_i)-\bar m_\mathrm{B}^{\rm ANN}(z_i)\big]
\big[m_\mathrm{B}^{(k)}(z_j)-\bar m_\mathrm{B}^{\rm ANN}(z_j)\big]\right\rangle.
\label{eq:CANN_def}
\end{equation}
This covariance captures the correlated uncertainty of the reconstructed SN curve induced by the global fit and by the correlated Pantheon+ errors.

On the FRB grid, the residual vector becomes
\begin{equation}
\mathbf{r}_\mathrm{B}(\theta,M_\mathrm{B})\equiv \bar{\mathbf{m}}_\mathrm{B}^{\rm ANN}-\mathbf{m}_\mathrm{B}^{\rm FRB}(\theta,M_\mathrm{B}).
\end{equation}
The total covariance now combines the ANN prediction covariance and the FRB reconstruction covariance evaluated on the same grid:
\begin{equation}
C_\mathrm{B} \equiv C_{\rm ANN}+C_{\rm FRB}.
\end{equation}
Equation~(\ref{eq:gauss_like_simple}) with $\mathbf{r}=\mathbf{r}_\mathrm{B}$ and $C=C_\mathrm{B}$ defines the ANN likelihood.


\subsection{Parametrizations of $\eta(z)$}
\label{sec:eta_param}

As the physical origin of a possible CDDR violation is a priori unknown (e.g.\ photon non-conservation, opacity, or more exotic effects), we adopt simple phenomenological one-parameter ans\"atze for
\(\eta(z)\equiv D_\mathrm{L}/[(1+z)^2D_\mathrm{A}]\) that are widely used in the distance-duality literature. We consider
\begin{align}
\text{linear:} \quad & \eta(z)=1+\epsilon z, \label{eq:eta_linear}\\
\text{linear fractional (linfrac):} \quad & \eta(z)=1+\epsilon\,\frac{z}{1+z}, \label{eq:eta_linfrac}\\
\text{power law (power):} \quad & \eta(z)=(1+z)^{\beta}, \label{eq:eta_power}
\end{align}
all satisfying \(\eta(0)=1\) by construction. The first two forms provide a minimal low-redshift expansion of a smooth deviation and are frequently adopted in CDDR tests because they capture the leading \(\mathcal{O}(z)\) behavior without committing to a specific microphysical model; the \(z/(1+z)\) choice further regularizes the growth at \(z\gtrsim 1\), avoiding an excessively steep extrapolation at the high-$z$ end of the sample. The power-law form offers an alternative, equally economical description and is also commonly employed in the context of cosmic-transparency/opacity-inspired CDDR tests, where the accumulated effect along the line of sight motivates simple scalings with \(1+z\). At low redshift, all three reduce to the same leading behavior, \(\eta(z)\simeq 1+\theta z+\mathcal{O}(z^2)\), with \(\theta=\epsilon\) (linear/linfrac) and \(\theta=\beta\) (power), so the parameters can be compared consistently at first order. In this work, we adopt flat priors $M_\mathrm{B}\in(-20.5,-18.5)$ and $\epsilon\in(-1,1)$ or $\beta\in(-2,2)$.

\begin{figure*}[htpb]
  \centering
  \includegraphics[width=\textwidth]{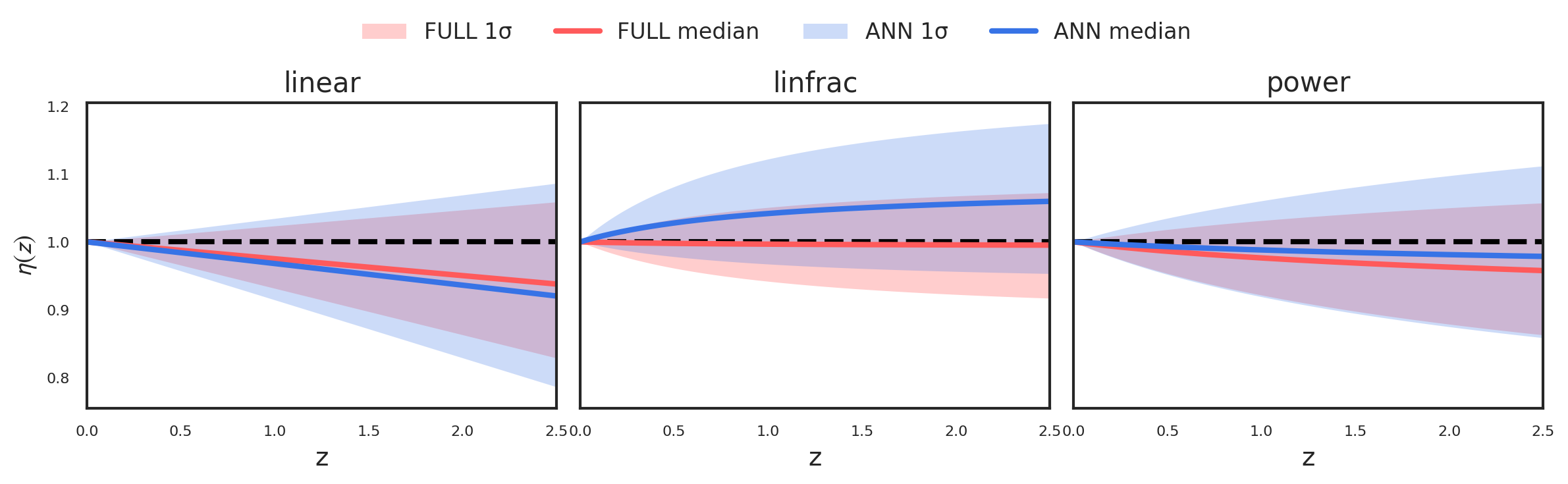}
  \caption{Reconstructed CDDR function $\eta(z)$ for the three parameterizations. Solid lines show posterior medians and shaded regions correspond to the 68\% (1$\sigma$) credible intervals. Black dashed line represents $\eta(z)=1$. Method A (FULL) and Method B (ANN) are compared, with results displayed in red and blue, respectively.}
  \label{fig:eta_panels}
\end{figure*}

\subsection{Pipeline overview}
\label{sec:pipeline_overview}

For clarity, we summarize here the end-to-end analysis pipeline used to compare $D_\mathrm{L}^{\rm SN}$ information with $D_\mathrm{A}^{\rm FRB}$ reconstruction:

\begin{itemize}
\item Build $\mathrm{DM}_{\rm EG}$ for each FRB by subtracting $\mathrm{DM}_{\rm MW}$ and $\mathrm{DM}_{\rm halo}$; apply quality cuts and remove DM outliers.
\item Reconstruct a smooth $ \overline{\mathrm{DM}}_{\rm EG}(z)$ with an ANN ensemble and bootstrap resampling to capture correlated uncertainties.
\item Anchor the mean trend at $z=0$ to enforce $ \overline{\mathrm{DM}}_{\rm IGM}(0)=0$, yielding an effective mean host/near-source term from the reconstructed intercept.
\item Obtain $H(z)$ from $\dd  \overline{\mathrm{DM}}_{\rm IGM}/\dd z$ (after removing the modulation $F(z)$), integrate to $D_\mathrm{C}(z)$, and compute $D_\mathrm{A}^{\rm FRB}(z)$; propagate the ensemble to a non-diagonal $C_{\rm FRB}$.
\item Use Pantheon+ corrected $m_\mathrm{B}$ and covariance; treat $M_\mathrm{B}$ as a nuisance parameter.
\item Constrain one-parameter $\eta(z)$ models with two likelihoods: Method A (direct SN with covariance submatrix) and Method B (SN reconstructed on FRB grid with non-diagonal $C_{\rm ANN}$).
\end{itemize}

\section{Results}\label{Sec7}

To place our analysis in context, it is worth emphasizing that CDDR tests increasingly rely on reconstruction-based comparisons to handle redshift mismatch between heterogeneous probes, rather than restricting to aggressively matched subsamples \cite{wang2024testing, li2025testing, luo2025testing, tang2023deep, dialektopoulos2023neural}. In this spirit, our two-likelihood strategy can be viewed as complementary implementations of the same underlying idea -- compare $D_\mathrm{L}^{\rm SN}$ tracer against $D_\mathrm{A}^{\rm FRB}$ baseline while properly propagating reconstruction-induced correlations across redshift. 

Table~\ref{tab:main} presents the posterior constraints on the CDDR-deviation parameter $\theta$ and $M_\mathrm{B}$ of SNe\,Ia for the three one-parameter families of $\eta(z)$ introduced in Sec.~\ref{sec:eta_param}. Results are shown for the two complementary likelihood implementations described in Sec.~\ref{sec:likelihood_method}: Method~A (FULL), which uses individual Pantheon+ supernovae with the corresponding covariance submatrix on the SN redshift subset, and Method~B (ANN), which reconstructs a smooth SN Hubble diagram on the FRB redshift grid while propagating the \emph{full} Pantheon+ covariance into a non-diagonal prediction covariance. 

In all cases, the recovered posteriors are statistically consistent with the Etherington relation, $\eta(z)=1$ (equivalently, $\theta=0$), and we find no evidence for a redshift-dependent violation of the distance duality relation at the current precision, indicating that the data do not prefer departures from the standard CDDR. This null result is also visualized in Fig.~\ref{fig:eta_panels}, where the reconstructed $\eta(z)$ bands (posterior medians with $1\sigma$ credible intervals) remain close to unity across the full redshift range probed by the localized FRBs. In particular, Method~A and Method~B yield mutually consistent $\eta(z)$ envelopes, which serves as a strong internal robustness check: despite the substantially different treatment of the SN information (direct vs.\ reconstructed), both approaches converge to the same physical conclusion.

To provide a more intuitive sense of scale, it is useful to translate the inferred $\theta$ values into an implied deviation of $\eta(z)$ at the high-redshift end. Taking the largest localized-FRB redshift in our sample, $z_{\max}\simeq 2.148$, the FULL constraints correspond to the following representative values: for the linear model, $\eta(z_{\max})\simeq 1+\epsilon z_{\max}\approx 0.95\pm0.10$; for the linfrac model, $\eta(z_{\max})\simeq 1+\epsilon z_{\max}/(1+z_{\max})\approx 0.995\pm0.076$; and for the power-law form, $\eta(z_{\max})\simeq (1+z_{\max})^\beta \approx 0.96\pm0.09$. These values (and their ANN counterparts) illustrate that the constraints correspond to at most $\approx10\%$-level allowed deviations at $z\sim 2$ given current data, with the posterior comfortably encompassing $\eta=1$. Importantly, the widening of the credible bands at $z\gtrsim 1$ in Fig.~\ref{fig:eta_panels} is expected, since the localized FRB sample becomes sparse at high redshift and correspondingly $D_\mathrm{A}^{\rm FRB}(z)$ broadens once propagated through the derivative-based pipeline described in Sec.~\ref{sec:dm_to_da}.

Among the three families, the linfrac parametrization naturally yields the weakest bounds on $\epsilon$ (and thus broader $\eta(z)$ bands) because its functional form saturates at high redshift: $z/(1+z)\to 1$ as $z\to\infty$. In practice, this reduces the leverage of the highest-$z$ objects relative to the linear model and makes the posterior more sensitive to the mid-$z$ region where the reconstruction covariances are non-trivial. Conversely, the linear and power-law models display similar constraining power in this dataset, as they both encode an approximately linear departure at low $z$ and accumulate a monotonic redshift trend that is more directly constrained by the global comparison between the SN and FRB distance ladders.

Another central methodological result is the consistency between Method~A and Method~B. These are deliberately different statistical constructions: Method~A preserves the SN data granularity, whereas Method~B introduces an additional reconstruction layer that inevitably broadens the posterior due to the finite reconstruction variance and smoothing-induced correlations. Consistent with this expectation, Method~B constraints are typically slightly weaker than Method~A, but always fully compatible within the quoted credible intervals (Table~\ref{tab:main}). This agreement indicates that our conclusions are not an artifact of a particular choice of data compression or interpolation scheme, and that the dominant inference is driven by the coherent redshift dependence encoded in the full covariance-aware comparison.

This consistency also provides a non-trivial cross-check of the covariance propagation. In Method~B, the SN reconstruction covariance (denoted $C_{\rm ANN}$ in Sec.~\ref{sec:methodB_new}) encodes the correlated uncertainty of the ANN-predicted curve induced by both the global nature of the fit and the underlying correlated Pantheon+ errors. In both methods, the FRB contribution enters as a non-diagonal covariance matrix $C_{\rm FRB}$, inherited from the bootstrap/ensemble reconstructions of $\langle \mathrm{DM}_{\rm EG}(z)\rangle$ and propagated through the derivative and integration operations required to obtain $D_\mathrm{A}^{\rm FRB}(z)$. Retaining these off-diagonal structures is essential: it prevents artificial over-tightening that would arise from treating neighboring redshift
points as independent when they are not.

\begin{figure}[htpb]
  \centering
  \includegraphics[width=0.48\textwidth]{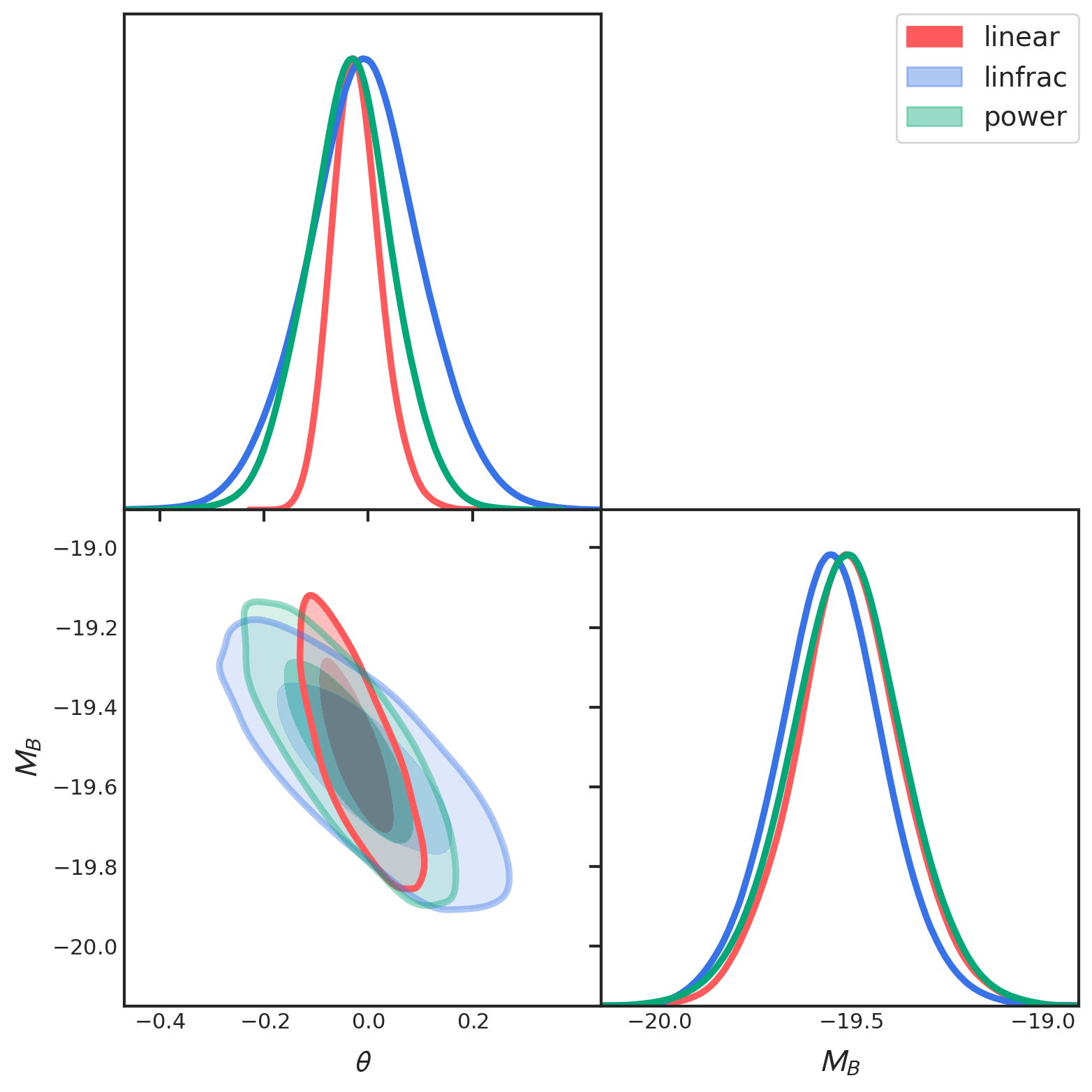}
  \caption{Posterior constraints in the $(\theta,M_\mathrm{B})$ plane for the three $\eta(z)$ families using Method A (FULL). Contours show 68\% (1$\sigma$) credible regions; $\theta=\epsilon$ for linear/linfrac and $\theta=\beta$ for power.}
  \label{fig:corner_full}
\end{figure}

\begin{figure}[htpb]
  \centering
  \includegraphics[width=0.48\textwidth]{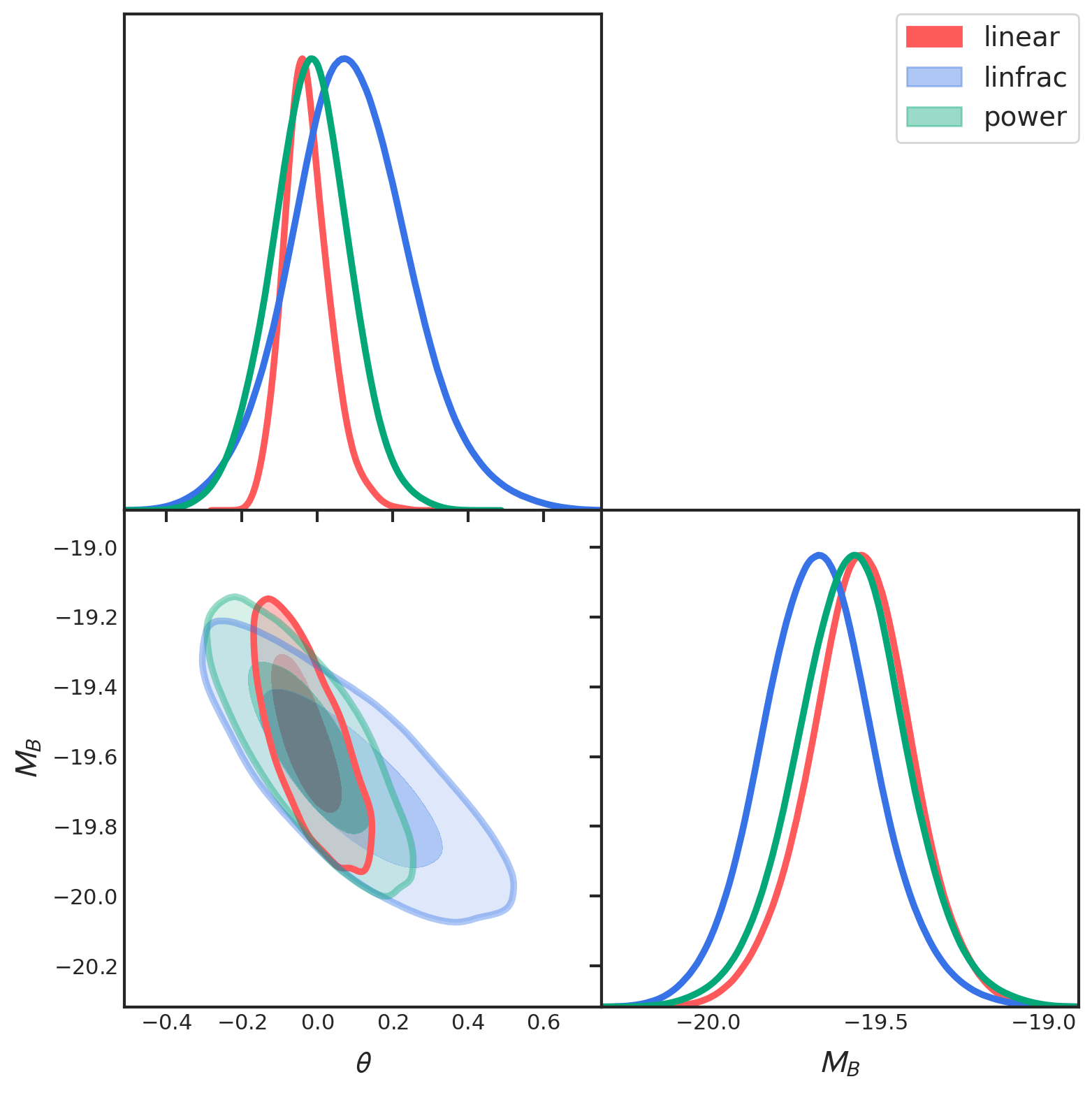}
  \caption{Same as Fig.~\ref{fig:corner_full}, but for Method B (ANN).}
  \label{fig:corner_ann}
\end{figure}

Figs.~\ref{fig:corner_full} and \ref{fig:corner_ann} show the joint posteriors in the $(\theta, M_\mathrm{B})$ plane. In all parametrizations, we observe a pronounced anti-correlation between $\theta$ and $M_\mathrm{B}$. This degeneracy has a clear physical origin. In magnitude space, $\eta(z)$ enters additively through $\Delta_\eta(z)=5\log_{10}\eta(z)$ [Eq.~\eqref{eq:delta_eta}], while $M_\mathrm{B}$ acts as a global offset. Over a finite redshift interval, a smooth redshift-dependent tilt in $\eta(z)$ can partially mimic an effective shift in the SN zero-point calibration. The data break this degeneracy through the \emph{shape} information of the Hubble diagram and through the redshift dependence induced by the FRB reconstruction, but the residual correlation remains visible in the contours.

The inferred $M_\mathrm{B}$ values are broadly stable across models and methods, typically in the range $M_\mathrm{B}\simeq -19.5$ to $-19.7$ with $\sim 0.15$--$0.17$\,mag uncertainties for the present analysis (Table~\ref{tab:main}). The mild shifts between Method~A and Method~B are consistent with the fact that Method~B uses a reconstructed mean SN curve on the FRB grid, effectively re-weighting information across redshift compared to the direct SN-by-SN likelihood. Importantly, these shifts do not translate into a preference for non-zero $\theta$, reinforcing that the conclusion $\eta(z)=1$ is robust to the treatment of the SN calibration nuisance.

\begin{figure*}[htpb]
  \centering
  \includegraphics[width=0.99\textwidth]{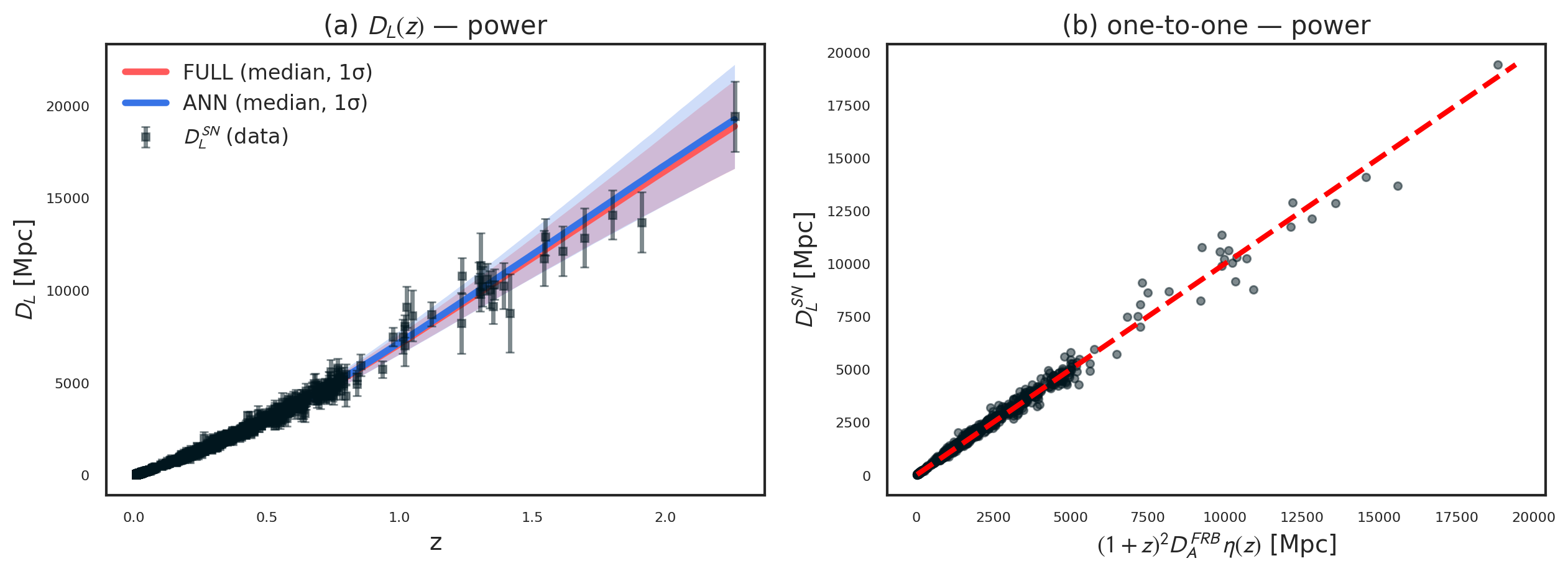}
  \caption{Distance-space diagnostics for the CDDR comparison (example shown for the \textit{power} model). The shaded bands show 68\% (1$\sigma$) credible intervals. Left: $D_\mathrm{L}(z)$ from SNe compared to $(1+z)^2D_\mathrm{A}^{\rm FRB}(z)\eta(z)$. Right: one-to-one comparison with the $y=x$ line.}
  \label{fig:dl_diag}
\end{figure*}

Beyond parameter constraints, Fig.~\ref{fig:dl_diag} provides a distance-space diagnostic of the pipeline. Using the posterior for $M_\mathrm{B}$, we map the SN magnitudes into an inferred $D_\mathrm{L}$ relation and compare it to the FRB-based prediction $(1+z)^2D_\mathrm{A}^{\rm FRB}(z)\eta(z)$ (shown for the power-law case). The overlap of the two bands across the full redshift interval indicates that $D_\mathrm{A}^\mathrm{FRB}$ reconstruction is consistent with $D_\mathrm{L}^\mathrm{SN}$ information once the calibration nuisance is marginalized over. In the one-to-one panel, points cluster around the $y=x$ relation within uncertainties, as expected for a null-duality-violation outcome.

As anticipated, the uncertainty in the FRB-based distance prediction broadens towards the highest redshifts. This is a direct reflection of (i) the decreasing density of localized FRBs at $z\gtrsim 1$ and (ii) the fact that our FRB distance inference is derivative-based, so the high-$z$ tail is more sensitive to the reconstruction variance of $\langle \mathrm{DM}_{\rm EG}(z)\rangle$ and its slope. Crucially, these effects are not treated heuristically: they are quantitatively captured by the ensemble-derived covariance $C_{\rm FRB}$ that enters both likelihoods.

\section{Discussions}\label{Sec8}

In this work, we have presented a cosmological model-independent test of the Etherington cosmic distance--duality relation by combining $D_\mathrm{L}$ information from SNe\,Ia with $D_\mathrm{A}$ baseline reconstructed from localized FRBs. The FRB side is particularly appealing for a distance-duality test because it is not based on photon fluxes: the inference is driven by the redshift evolution of the extragalactic dispersion measure and by the geometric mapping from the reconstructed expansion history to $D_\mathrm{A}(z)$. Within this framework, we find no evidence for a redshift-dependent violation of the CDDR, and the reconstructed $\eta(z)$ remains compatible with unity across the redshift interval probed.

A key aspect of our approach is that it avoids imposing a specific parametric cosmological model for $H(z)$ or for the distance scale. Instead, the expansion history is recovered from a smooth reconstruction of the mean extragalactic dispersion measure component from FRBs. This makes the test a genuine consistency check: any detected redshift trend in $\eta(z)$ would be difficult to attribute to the usual cosmology-prior assumptions, and would more directly point to photon non-conservation, opacity-like effects, non-flat spatial cosmic curvature, or residual systematics in one of the distance probes. 

On the FRB side, our reconstruction strategy also provides a useful physical anchor at very low redshift. By enforcing $ \overline{\mathrm{DM}}_{\rm IGM}(0)=0$ at the level of the reconstructed mean, we effectively isolate an intercept term $ \overline{\mathrm{DM}}_{\rm EG}(0)$ that can be interpreted as the average host contribution. This is not an external prior, but a consistency requirement of the dispersion-measure budget: the cosmological IGM contribution must vanish as $z\rightarrow 0$. In practice, this anchoring mitigates unphysical behavior in the low-$z$ tail and yields a direct, data-driven constraint on the effective mean host term that is propagated through the pipeline.

Finally, the statistical framework developed here is designed to be robust to reconstruction-induced correlations. Because both the FRB distances and (in Method~B) the SN Hubble diagram are obtained through global reconstructions, the resulting uncertainties are correlated across redshift; treating them as independent would bias the inference by artificially tightening constraints. By retaining the full non-diagonal covariance structure in both methods, we ensure that the CDDR constraints are driven by coherent redshift-dependent information rather than by over-counting effectively smoothed data points. This covariance-aware, model-independent strategy is directly extensible to future localized FRB samples and provides a clean route to progressively sharper CDDR tests as FRB redshift completeness and localization statistics continue to improve.\\


\acknowledgments

JASF thanks the National Research Foundation (NRF) of South Africa for their financial support. SK and AW gratefully acknowledge the support and hospitality of the ICTP, Italy and the Simons Fellow programme. This work was developed during a research visit supported by the Simons Associates Programme. SK acknowledge funding from the National Science Centre, Poland (grant no. 2023/49/B/ST9/02777). We also gratefully acknowledge supportfrom the South African Research Chairs Initiative of the Department of Science and Technology
and the National Research Foundation. 

\bibliographystyle{apsrev4-1}
\bibliography{frb_bib.bib}

\end{document}